\documentstyle[amssymb,preprint,aps,epsf]{revtex}
\global\firstfigfalse

\begin{document}
\title{A Single Impurity in Tomonaga--Luttinger Liquids.}
\author{Yuval Oreg and Alexander M. Finkel'stein\cite{Lan}}
\address{Department of Condensed Matter Physics, The Weizmann\\
Institute of Science,\\
Rehovot 76100, Israel}
\date{March 1997}
\maketitle

\begin{abstract}
\label{abstract}

The problem of a single impurity in one dimensional Tomonaga --Luttinger
liquids with a repulsive electron--electron interaction is discussed. We
find that the renormalization group flow diagram for the parameters
characterizing the impurity is rather complex. Apart from the fixed points
corresponding to two weakly connected semi--infinite wires, the flow diagram
contains additional fixed points which control the low temperature physics
when the bare potential of the impurity is not strong.
\end{abstract}

%\maketitle

%\pacs{PACS numbers: 78.70.Dm, 79.60.Jv, 72.10.Fk}

\makeatletter
\global\@specialpagefalse

\let\@evenhead\@oddhead
\makeatother

\section{Introduction}

\label{intro}

The recent advance in submicron technology enables fabrication of truly
one--dimensional (1D) quantum wires. The electron liquids in these systems
are usually described in terms of the Tomonaga--Luttinger (TL) model\cite
{RFS:Mahan90,RFS:Mattis93}. Edge states in a two--dimensional electron gas,
under conditions of the fractional quantum Hall effect, were argued~to be TL
liquids as well\cite{FQHE:Wen91}. It is well known that in the TL\ model
with a repulsive electron--electron interaction the effective strength of a
backward scattering by an impurity defect increases with decreasing
temperature\cite{EE1D:Mattis74b}. For this reason, the conductance of TL
liquids with a single defect has been intensively discussed recently using
various theoretical methods\cite
{EE1D:Kane92a,EE1D:Matveev93,1DT:Furusaki93,1DT:Ogata94,EE1D:Fendley95}. It
has been concluded that even a weak impurity eventually causes the
conductance to vanish at low temperatures.

The physical interpretation of the conductance vanishing is based on the
assumption{\em \ }made by Kane and Fisher\cite{EE1D:Kane92}, that in the
limit of low temperatures the behavior of a TL system with an impurity may
be described as tunneling between two disconnected semi-infinite TL wires.
The effective amplitude of tunneling between the half--wires scales to zero
with decreasing temperature, because the tunneling density of states at the
ending point of a TL liquid vanishes when the electron--electron interaction
is repulsive\cite{EE1D:Kane92,EE1D:Matveev93a}. This interpretation
corresponds to a scenario in which the effective strength of the impurity
increases in the course of the renormalization, so that at the final stage a
weak impurity transforms into a strong barrier, and disconnects the TL wire.
However, a direct calculation of the tunneling density of states\cite
{EE1D:Oreg96}, obtained by a mapping of the weak impurity problem onto a
Coulomb gas theory, apparently contradicts this intuitive\ picture. It has
been found that at the location of a weak impurity the tunneling density of
states is enhanced, rather than vanishing. The scenario of Ref.\cite
{EE1D:Kane92} is based on the assumption that no other fixed points
intervene in the scaling from the repulsive fixed point of a weakly
scattering defect to the attractive fixed point corresponding to a tunneling
junction of two half-wires. The contradiction of this scenario to the
calculations of single particle properties, such as the tunneling density of
states, and the Fermi edge singularity\cite{XR:Oreg96}, indicates that maybe
this is not the case.

In this work the problem of a single impurity in TL liquids with a repulsive
electron--electron interaction is reinvestigated. We concentrate on the
limit when the Fermi wave length is much larger than the defect size. This
situation is typical for semiconductors, where the filling of the conduction
band is far from one half. Then it is possible to describe the problem as a
continuous model with an appropriately chosen point--like defect. We come to
the conclusion that the low energy physics of a weak impurity is controlled
by a fixed point that differs from the one corresponding to a tunneling
junction of two half-wires. This aspect of the discussed problem proves to
be similar to an overscreened two channel Kondo~problem, while the original
idea of the theory of Ref.\cite{EE1D:Kane92} has been orientated to a
situation similar to the ordinary Kondo problem (see also Ref.\cite
{RG1D:Eggert92} for a discussion). The above statement may seem to be
strange. Indeed, in the overscreened two channel Kondo~problem the
infinitely strong exchange interaction corresponds to a repulsive fixed point%
\cite{RG1D:Nozieres80}, while in the problem under discussion the only
apparent candidate for a fixed point in the strong coupling regime is the
fixed point of a tunneling junction\cite{RG1D:Weaklink}, which is
attractive. We find here that there exists another fixed point corresponding
to the infinitely strong amplitude of the impurity backward scattering, and
it is repulsive. This implies that in addition there should be an attractive
fixed point at a finite value of the backward scattering amplitude. The
renormalization group (RG) flow diagram of the problem depends on several
parameters characterizing the impurity, and is rather complex. Each of the
fixed points describing the backward scattering problem, as well as that of
the tunneling junction, corresponds to a manifold of fixed points. An
important fact is, that the tunneling junction manifold, and the two of the
new fixed points describing backward scattering, are located in different
parts of the RG flow diagram.

For a point--like defect, all Fourier components of an impurity potential
are practically the same. Therefore, one may characterize the ''strength''
of the potential by an initial (bare) value of a dimensionless parameter $%
u_{+}=\int U(x)dx/v_{F},$ where $v_{F}$ is the Fermi velocity. When $u_{+}$
is small, the backward scattering problem is described by the partition
function $Z_{sc}(u_{-};g)$, depending on a backward scattering amplitude $%
u_{-}$, and on the electron--electron interaction parameter $g$.
Alternatively, when $u_{+}$ is large, i.e., the impurity is strong, the
appropriate description is a tunneling model with a partition function $%
Z_{tun}(t_{-};g)$, depending on a tunneling amplitude $t_{-}$. For free
electrons, the two partition functions are deeply connected, because of the
relation between the reflection from and the transmission through the
defect. In this case, in principle, each one of them can represent an
impurity of an arbitrary strength. The situation becomes quite different
when the electron--electron interaction is switched on in TL liquids. Then,
the partition functions $Z_{sc}(u_{-};g)$ and $Z_{tun}(t_{-};g)$\ are not
equivalent anymore. Instead, they describe two complementary limiting cases
of the phase diagram, and give completely different structures of the RG
fixed points for these cases.

In this paper, the RG flow diagram of the continuous model is analyzed by
examining different ''corners'' of the scattering and the tunneling limits,
and tailoring them together. The flow diagram is constructed in axes
representing the strength of the impurity $u_{+}$, and the backward
scattering amplitude $u_{-}$ (and/or $1/t_{-}$), respectively. The
presentation of the phase diagram in the space of the impurity potential
parameters, rather than the conventional presentation in terms of the
backward scattering amplitude together with the parameter $g$, helps clarify
the different character of the fixed points controlling the low energy
physics in the cases of a weak and a strong bare impurity potential. In the
course of renormalization of a weak impurity the effective barrier evolves
in a special manner. Only the backward scattering amplitude $u_{-}$
increases, while $u_{+}$, being marginal, remains small. That is why a weak
local impurity in a continuous model does not evolve towards a strong
barrier, but proceeds to scale in the other, complementary, part of the RG
flow diagram. The low energy physics in this case is controlled by a line of
fixed points $L_{f}$, at a finite value of $u_{-}$. (More precisely, $L_{f}$
is a manifold of fixed points due to the presence of other marginal
parameters. Since we use a two--dimensional plot for the RG-diagram, we will
call such manifolds ''lines''.) The other part of the flow diagram contains
a line of attractive fixed points, $D_{0}$, controlling the low energy
physics when the bare potential of the impurity is strong. This line
describes the vanishing of the tunneling amplitude $t_{-}$, and corresponds
to two disconnected half-wires.

We think that the attraction of the weak impurity problem to $L_{f}$, but
not to $D_{0}$, is the reason for the difference between the results
obtained by means of the Coulomb gas theory for the single particle
correlators, and those obtained relying on the scenario of two disconnected
wires.

The paper is organized as follows: in Sec.~\ref{se:weakstrong} we discuss
the renormalization of the scattering and the tunneling Hamiltonians, i.e.,
the RG flow for two different limits of the impurity strength. The existence
of several parameters describing a local defect in the continuous model is
emphasized. In Sec.~\ref{se:RGphaseplane} a RG flow diagram unifying both
cases of a weak and of a strong impurity potential is discussed. In Appendix~%
\ref{app1} the repulsive character of the limit $u_{-}=\infty $ and,
consequently, the existence of the attractive $L_{f}$-line, is demonstrated
by mapping the impurity problem onto a spin-$1/2$ Heisenberg chain.

\section{The regions of a weak and a strong impurity in the flow diagram}

\label{se:weakstrong}

In this section we reproduce the main results concerning the renormalization
of the problem in the limits of a weak and of a strong impurity potential.
We assume that the Fermi wave length is much larger than the defect size.
This situation occurs in semiconductor wires where the filling of the
conduction band is far from one half and the Fermi wave length is large.
Several authors used the half-filled tight binding model with a link defect,
to analyze the problem of backward scattering in quantum wires. The tight
binding model corresponds to a fixed choice of the parameters describing the
defect, that makes this model a specific one. In this case the Fermi wave
length is commensurate with the impurity size, and the final fixed point
depends on the internal structure of the defect\cite{RG1D:Eggert92}. The
continuous model considered here is not constrained by the specific
structure of the local defect.

\begin{center}
{\bf A weak potential scattering: the repulsive line }$L_{0}$
\end{center}

The \ Hamiltonian of the TL model is given by

\begin{equation}
H_{TL}=-iv_{F}\int dx\psi _{R}^{\dagger }(x)\frac{\partial }{\partial x}\psi
_{R}(x)+iv_{F}\int dx\psi _{L}^{\dagger }(x)\frac{\partial }{\partial x}\psi
_{L}(x)+\frac{1}{2}V\int dx\left( \rho _{R}(x)+\rho _{L}(x)\right) ^{2},
\label{rgeq:ffh}
\end{equation}
Here the electron spectrum is linearized near the Fermi points $\pm k_{F}$; $%
\psi _{R}(x)\mbox  {  and  }\psi _{R}^{\dagger }(x)$ are the field operators
of fermions that propagate to the right with wave vectors $\approx +k_{F}$,
and $\psi _{L}(x)\mbox { and }\psi _{L}^{\dagger }(x)$ are the field
operators of left propagating fermions with wave vectors $\approx -k_{F}$; $%
\;\rho _{L\left( R\right) }(x)=\psi _{L(R)}^{\dagger }(x)\psi _{L\left(
R\right) }(x)$ are the electron density operators; $V$ describes the
density--density interaction with a momentum transfer much smaller than $%
k_{F}$. The Hamiltonian (\ref{rgeq:ffh}) describes the 1D electron liquid
when the backward scattering amplitude of the electron--electron interaction
may be ignored, and it is a fixed point Hamiltonian for a broad class of 1D
systems.

For low energy physics only processes of electron scattering with a momentum
transfer close to zero and to $2k_{F}$ are essential. Let us denote 
\begin{equation}
U_{0}=\int dxU(x),\;\;\;\;\;U_{2k_{F}}=\int dxU(x)e^{i2k_{F}x}.
\label{eq:def}
\end{equation}
Here $U_{0}$ and $U_{2k_{F}}=-|U_{2k_{F}}|e^{i\varphi _{u}}$ are the Fourier
transform amplitudes of the impurity potential $U(x).$ For a weak impurity
the forward and backward scattering amplitudes are

\begin{equation}
u_{+}=U_{0}/v_{F},\qquad u_{-}=|U_{2k_{F}}|/v_{F},
\end{equation}
respectively; $v_{F}$ is the Fermi velocity. (To avoid confusion, we will
not use here the notations $\delta _{\pm }$ for the dimensionless amplitudes 
$u_{\pm }$, unlike in \cite{XR:Oreg96}, because the relations of the
amplitudes $u_{\pm }$ with the scattering phase shifts are ill defined for
interacting electrons in the TL model.) Note, that since the bare impurity
potential $U(x)$ is assumed to be local, it has Fourier components of the
same order for practically all momenta. Therefore, the line of the bare
parameters corresponds to $u_{+}\approx u_{-}$.

In addition to $u_{+}$, $u_{-}$, and $\varphi _{u}$, another parameter, $%
u_{a}$, describing the asymmetry of the forward scattering of left and right
movers will be introduced. In the presence of time reversal symmetry $%
u_{a}\equiv 0$, but it is not necessarily zero for the quantum Hall edge
states. The scattering of the conduction band electrons by a single local
defect at $x=0$ is given by

\begin{equation}
\begin{array}{lll}
H_{sc} & = & v_{F}\left[ u_{+}\left( \psi _{R}^{\dagger }(0)\psi
_{R}(0)+\psi _{L}^{\dagger }(0)\psi _{L}(0)\right) +u_{a}\left( \psi
_{R}^{\dagger }(0)\psi _{R}(0)-\psi _{L}^{\dagger }(0)\psi _{L}(0)\right)
\right. \\ 
&  & -\left. u_{-}\left( e^{i\varphi _{u}}\psi _{R}^{\dagger }(0)\psi
_{L}(0)+e^{-i\varphi _{u}}\psi _{L}^{\dagger }(0)\psi _{R}(0)\right) \right]
,
\end{array}
\label{rgeq:fih}
\end{equation}
where $\psi _{R(L)}(0)\equiv \psi _{R(L)}(x=0)$.

The bosonization technique (for review see \cite{RFS:Mahan90,RFS:Mattis93})
allows one to reduce the TL Hamiltonian to a quadratic form in terms of
operators of bosonic fields $\phi _{0}$ and $\tilde{\phi}_{0}$:

\begin{eqnarray}
\phi _{0}(x) &=&\frac{-i}{\sqrt{4\pi }}\frac{2\pi }{L}\sum_{p}\frac{\exp
\left( {-\eta |p|/2-ipx}\right) }{p}\left[ \rho _{R}(p)+\rho _{L}(p)\right] ,
\label{rgeq:dpd} \\
\tilde{\phi}_{0}(x) &=&\frac{-i}{\sqrt{4\pi }}\frac{2\pi }{L}\sum_{p}\frac{%
\exp \left( {-\eta |p|/2-ipx}\right) }{p}\left[ \rho _{R}(p)-\rho
_{L}(p)\right] ,
\end{eqnarray}
where $L$ is the system length, and $\eta ^{-1}$ is an ultraviolet cutoff, $%
\left( 2\pi \eta \right) ^{-1}$ is equal to the average density of the
electron liquid. The fields $\phi _{0}$ and its dual partner $\tilde{\phi}%
_{0}$ are conjugate variables, i.e., 
\begin{equation}
\left[ \frac{d\phi _{0}(x)}{dx},\tilde{\phi}_{0}(y)\right] =i\delta \left(
x-y\right) .  \label{rgeq:cr}
\end{equation}
After rescaling the operators 
\begin{equation}
\phi =\frac{\sqrt{4\pi }}{\beta }\phi _{0},\;\tilde{\phi}=\frac{\beta }{%
\sqrt{4\pi }}\tilde{\phi}_{0},  \label{rgeq:rphi}
\end{equation}
the bosonized representation of the Hamiltonian (\ref{rgeq:ffh}) becomes 
\begin{equation}
H_{TL}^{B}=\frac{v_{F}}{2g}\int dx\left( \left( \frac{d\phi }{dx}\right)
^{2}+\left( \frac{d\tilde{\phi}}{dx}\right) ^{2}\right) .  \label{rgeq:bfh}
\end{equation}
Here $g$ is an effective parameter of the electron--electron interaction 
\begin{equation}
g=\frac{\beta ^{2}}{4\pi },\;\beta ^{2}=4\pi \sqrt{\frac{1-\gamma }{1+\gamma 
}},\qquad \gamma =\frac{V}{\left( 2\pi v_{F}+V\right) };  \label{rgeq:abc}
\end{equation}
$g<1$ when the electron--electron interaction is repulsive, while for the
attractive interaction $g>1$.

The bosonic representations of the operators $\psi _{R}$ and $\psi _{L}$ are
given as 
\begin{eqnarray}
\psi _{R}^{B}(x) &=&\frac{e^{ik_{F}x}}{\sqrt{2\pi \eta }}\exp {\left[ -\frac{%
i}{2}\left( \frac{4\pi }{\beta }\tilde{\phi}+\beta \phi \right) \right] ,}
\label{rgeq:fermion} \\
\psi _{L}^{B}(x) &=&i\frac{e^{-ik_{F}x}}{\sqrt{2\pi \eta }}\exp {\left[ -%
\frac{i}{2}\left( \frac{4\pi }{\beta }\tilde{\phi}-\beta \phi \right)
\right] }.
\end{eqnarray}
Then, the scattering by a weak impurity may be written in terms of the $\phi 
$-field as

\begin{equation}
H_{sc}^{B}=\frac{v_{F}}{2\pi }\left( \beta u_{+}\left. \frac{d\phi }{dx}
\right| _{x=0}+\frac{4\pi }{\beta }u_{a}\left. \frac{d\tilde{\phi}}{dx}
\right| _{x=0}-\frac{2u_{-}}{\eta }\cos (\beta \phi (0)+\varphi _{u})\right)
.  \label{rgeq:bih}
\end{equation}
The RG equations for this problem are

\begin{mathletters}

\begin{equation}
\frac{du_{-}}{d\xi }=u_{-}(1-g),  \label{eq:rga}
\end{equation}

\begin{equation}
\frac{du_{+}}{d\xi }=0,  \label{eq:rgb}
\end{equation}
\begin{equation}
\frac{du_{a}}{d\xi }=0,\qquad \frac{d\varphi _{u}}{d\xi }=0,  \label{eq:rgc}
\end{equation}

\begin{equation}
\frac{dg}{d\xi }=0.  \label{eq:rgd}
\end{equation}

\end{mathletters}

The last equation here reflects the well known fact that in the TL model the
interaction parameter $g$ is not renormalized. Since $g$ describes a 'bulk'
electron liquid, the presence of a single impurity cannot influence this
parameter. The decoupling of the RG-equations for amplitudes $u_{+}$ from
the rest of the parameters is a consequence of an important property of the
Hamiltonian $H_{TL}+H_{sc}$. Namely, it can be split into two commuting
parts containing the forward $u_{+}$, and the backward $u_{-}$, scattering
terms separately\cite{XR:Kane94,XR:Affleck94}, see Eq.~(\ref{eq:LAH}). Since
the forward scattering $u_{+}$ is given by a linear term in Eq.~(\ref
{rgeq:bih}), and the amplitude $u_{-}$ cannot influence the renormalization
of $u_{+}$ because of the separation, the fulfillment of Eq.~(\ref{eq:rgb})
becomes evident. Hence, under the condition that the impurity can be
described as a local weak potential in a TL liquid, the forward scattering
amplitude $u_{+}$ is a marginal parameter. The parameters $u_{a}$ and $%
\varphi _{u}$ are marginal as well, see Eq.~(\ref{eq:rgc}). This is because
the $u_{a}$-term, being linear in $d\tilde{\phi}/dx$, can be removed from
the Hamiltonian by a canonical transformation, while $\varphi _{u}$ does not
reveal itself in the partition function $Z_{sc}$. In contrast, the backward
scattering amplitude is relevant for the repulsive case. Eq.~(\ref{eq:rga}%
)~has been derived \cite{EE1D:Kane92a} (see also \cite{EE1D:Bulgadaev82})
for small $u_{-}$. The renormalization of $u_{-}$ in the strong coupling
regime is discussed in Sec.~\ref{se:RGphaseplane}.

\begin{figure}[htb]
\epsfysize=6cm
\begin{center}
\leavevmode
\epsfbox{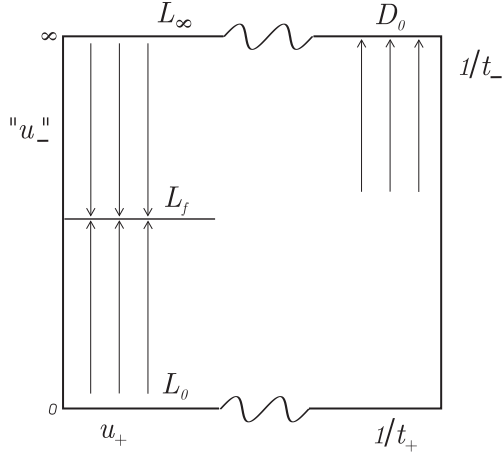}
\end{center}
\caption{{\em The RG-flow diagram of a point-like defect in a quantum wire
with a repulsive electron--electron interaction:} The line $L_{0}$
represents a manifold of fixed points in the case of an infinitesimally
small impurity potential. The line $D_{0}$ corresponds to the tunneling
junction limit. The line $L_{\infty}$ corresponds to the limit of an
infinitely large backward scattering amplitude $u_{-}$ in Hamiltonian~(\protect{\ref
{rgeq:bih}}). The attractive fixed line $L_{f}$ controls the low temperature
physics of the backward scattering problem. The problems of a weak impurity
and a weak tunneling junction evolve at complementary parts of the phase
diagram, and do not flow to each other in the model with a linearized
electron spectrum. }
\label{fg:RGTL}
\end{figure}

Eqs.~(\ref{eq:rga}--\ref{eq:rgd}) describe a repulsive manifold of fixed
points $L_{0}$, denoted as the $L_{0}$-line at the left-bottom corner, $%
u_{\pm }\ll 1$, of the RG-plane depicted in Fig.~\ref{fg:RGTL}.

\begin{center}
{\bf A strong potential barrier: the line of attractive fixed points }$D_{0}$
\end{center}

When the bare impurity is strong enough, the description of the problem in
terms of the scattering amplitudes $u_{-},u_{+},u_{a}$, and $\varphi _{u}$
ceases to be adequate. On the other hand, the fact that the impurity is
strong does not contradict the assumption of locality $k_{F}a\ll 1$. A local
and strong impurity can also be considered as a point--like problem, but
here it should be described by two semi--infinite TL liquids, with a weak
tunneling junction between their ending points. Like in the case of the weak
potential scattering, there are four parameters that describe the tunneling
and reflection processes at the tunneling junction of the two half--wires.
These parameters are the tunneling amplitude $t_{-}$, its phase $\varphi
_{t} $, and the two parameters, $t_{+}$ and $t_{a}$, characterizing the
phases that an electron acquires when it is reflected at the ends of the
half--wires. The parameter $t_{a}$ describes the asymmetry of the left and
the right parts of the tunneling junction. In the particular case of a
strong $\delta $-function potential, where $u_{\pm }=u$ $\gg 1$, the
amplitudes $t_{\pm }$ $\sim 1/u$.

The low energy physics of each semi--infinite wire may be described by a
single{\em \ }chiral mode\cite{EE1D:Kane92a,XR:Affleck94}. It will be
assumed that there is no density--density interaction between the
half--wires, but inside each of them the density--density interaction is
present. The effects of the electron--electron interaction inside a
half--wire can be taken into consideration by a canonical transformation
(see ,e.g., Sec. ${\em IIIA}$ in Ref.\cite{XR:Oreg96}). Then the appropriate
bosonic description of tunneling between two oppositely moving chiral modes
corresponding to each of the half-wires is given by 
\begin{equation}
H_{t}^{B}=\frac{v_{F}}{2\pi }\left( \frac{4\pi }{\beta }t_{+}\left. \frac{%
d\phi _{tun}}{dx}\right| _{x=0}+\beta t_{a}\left. \frac{d\tilde{\phi}_{tun}}{%
dx}\right| _{x=0}-\frac{2t_{-}}{\eta }\cos \left( \frac{4\pi }{\beta }\phi
_{tun}(0)+\varphi _{t}\right) \right) .  \label{eq:Ht}
\end{equation}
Here $d\phi _{tun}(x)/dx$ is related to the total density of the two chiral
modes, and $\tilde{\phi}_{tun}$ is a field dual to $\phi _{tun}$.

The RG equations in this case are analogous to Eqs.~(\ref{eq:rga}-\ref
{eq:rgd}):

\begin{mathletters}

\begin{equation}
\frac{dt_{-}}{d\xi }=t_{-}(1-1/g),  \label{eq:rgta}
\end{equation}

\begin{equation}
\frac{dt_{+}}{d\xi }=0,  \label{eq:rgtb}
\end{equation}
\begin{equation}
\frac{dt_{a}}{d\xi }=0,\qquad \frac{d\varphi _{t}}{d\xi }=0.  \label{eq:rgtc}
\end{equation}

\end{mathletters}

In contrast to $u_{-}$, the tunneling amplitude $t_{-}$ scales to zero for
repulsive electron--electron interaction ($g<1$). Therefore, the manifold of
fixed points described by the tunneling Hamiltonian $H_{t}^{B}$ is
attractive. In the two--dimensional plot depicted in Fig.~\ref{fg:RGTL} it
is presented in the upper right corner as the $D_{0}$ -line.

\section{ The unified RG flow diagram}

\label{se:RGphaseplane} The problem of an impurity in a TL liquid has been
discussed in Sec.~\ref{se:weakstrong} for the two limiting models: an
impurity scattering, Eq.(\ref{rgeq:bih}), and a tunneling junction, Eq.(\ref
{eq:Ht}). The scaling flow diagrams of these complementary models are
depicted together in Fig.~\ref{fg:RGTL} as a combined plot. In this plot,
the scattering model is represented on the left side, and the tunneling
model -- on the right one. The center of the plot corresponds to the
crossover region when the impurity potential has an intermediate strength.
Since for a large barrier the backscattering amplitude is large, and the
tunneling amplitude is small, we use the vertical axis to represent $u_{-}$
and $1/t_{-}$. The horizontal axis represents $u_{+}$ together with the
other parameters, which in a model with a linearized electron spectrum are
not renormalized. Due to the presence of these additional parameters, lines
on the symbolic two-dimensional plot correspond to hyperplanes.

For a point-like impurity there are certain relations between the bare
parameters characterizing the potential: $u_{-}\approx u_{+}$, and $%
1/t_{-}\approx 1/t_{+}$. Hence, only the left-bottom and the right-upper
corners of the phase diagram in Fig.~\ref{fg:RGTL} are attainable for bare
parameters of a local impurity in the continuous model. However, according
to Eqs.~(\ref{eq:rga}), (\ref{eq:rgb}) the amplitude $u_{-}$ renormalizes to
a strong coupling regime, while $u_{+}$ being marginal, remains small.
Therefore, to understand the nature of this regime, we have to analyze the
scattering Hamiltonian (\ref{rgeq:bih}) for the region of the parameters
corresponding to the upper left corner of the phase diagram. (Since large $%
u_{-}$ together with small $u_{+}$ does not correspond to a physical
realization of the bare parameters, we use in the phase diagram ``$u_{-}$''
with quotation marks.) In Appendix~\ref{app1} , the line $L_{\infty }$ with $%
u_{-}\gg 1$ and a not too large $u_{+}$ is argued to be a repulsive fixed
line. For this purpose, the Hamiltonian $H=H_{TL}+H_{sc}$ of the impurity
scattering in the TL model is mapped onto a semi-infinite spin-$1/2$
Heisenberg chain. In this mapping the impurity backscattering amplitude
corresponds to a magnetic field $h\propto u_{-}$, that acts on the spin
located at the origin of the chain. In the course of the RG treatment small $%
h$ scales, as $u_{-}$ does, to a strong coupling regime. To investigate the
nature of this regime, we analyze the stability of the limit $h=\infty $,
following the spirit of the treatment of the two channel Kondo problem by
Nozi\`{e}res and Blandin\cite{RG1D:Nozieres80}. When $h$ is very large, the
spin--spin coupling in the spin chain can be considered as a small
perturbation in a real space RG analysis. This procedure generates a new
operator, which is not irrelevant. If one assumes that $h\rightarrow \infty $
is a stable fixed point, the latter fact makes the RG process nonconvergent.
This is in contradiction to the assumption of stability, and therefore we
have to conclude that $h\rightarrow \infty $ is a repulsive fixed point.
Since $h\propto u_{-}$, we get that the limit of an infinitely strong
backward scattering is repulsive. Both limits, $u_{-}=0$ and $u_{-}=\infty $%
, are repulsive, and therefore there necessarily should be an attractive
fixed point at some finite value of the backward scattering amplitude $u_{-}$%
. (This analysis is valid when a description of the local impurity
scattering in terms of the TL model is a good approximation. Then the
Hamiltonian $H$ can be split into two commuting parts describing the
backward and the forward scattering separately\cite{XR:Kane94,XR:Affleck94},
see Eq.~(\ref{eq:LAH}). The discussed RG-regime of the $u_{-}$-amplitude
develops inside the backward scattering part alone.){\em \ }Thus, on the
left side of the diagram there are two repulsive lines $L_{0}$ and $%
L_{\infty }$, and in addition an attractive line $L_{f}$, which is
sandwiched between the opposite directions of the flow. The right side of
the diagram corresponding to the tunneling Hamiltonian (\ref{eq:Ht}),
contains an attractive fixed line $D_{0}$ which describes scaling to zero of
the tunneling amplitude $t_{-}$.

The information collected up to now is presented on the combined plot
in Fig.~\ref{fg:RGTL}. To complete the central part of the flow
diagram, a region of an intermediate impurity strength should be
studied. None of the two limiting models describes the problem
faithfully in this crossover region, and a consideration of a more
comprehensive Hamiltonian, which covers both limiting cases, is
needed. Moreover, to study the RG-flow in the crossover region, one
has to give up the approximations of the linearized electron spectrum
and/or of the locality of the defect. Then, the decoupling of forward
and backward scattering is no longer valid. (In the bosonized
representation terms $\propto \rho_{L(R)}^3$ describe the curvature
of the electron spectrum. One of the possible ways to consider the
nonlocality of the impurity is to add a term $\propto
\psi^{\dagger}_R(0) \frac{d}{dx} \psi_L(0) +h.c.$. When $k_Fa \ll 1$
the coefficient of this term is very small.)  Effects arising due to
the nonlinearity of the electron spectrum and the nonlocality of the
defect should be studied by a loop expansion in higher orders.  These
effects may have highly important influence on the renormalization of
the parameters $u_{+}$ and $u_{-}$. For example, it may cause the flow
lines near the $L_{f}$-line to bend to the left or to the right.

\begin{figure}[htb]
\hskip 1cm
\epsfysize=5cm
\leavevmode
\epsfbox{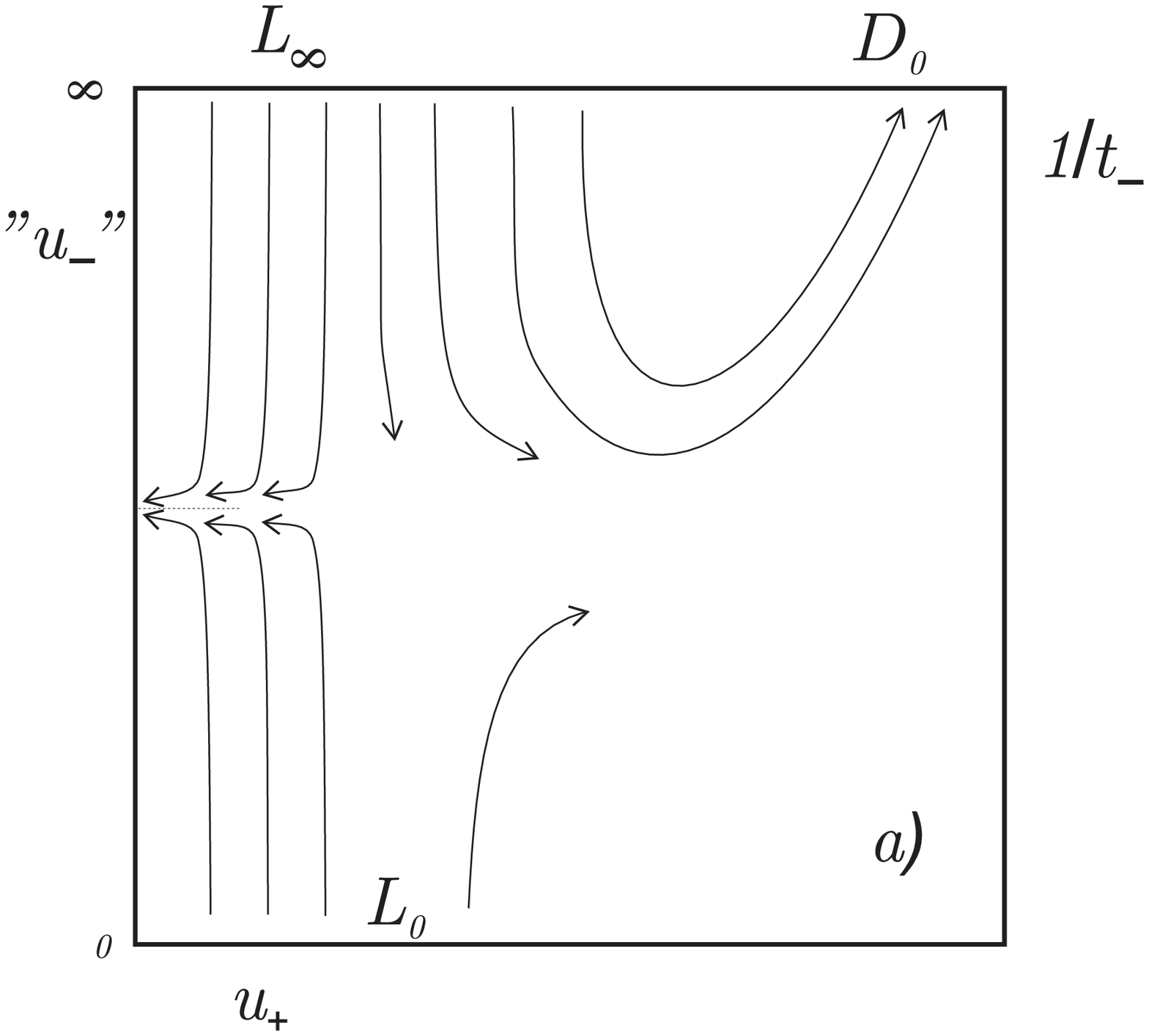}
\vskip -5cm
\hskip 7.5cm
\epsfysize=5cm
\leavevmode
\epsfbox{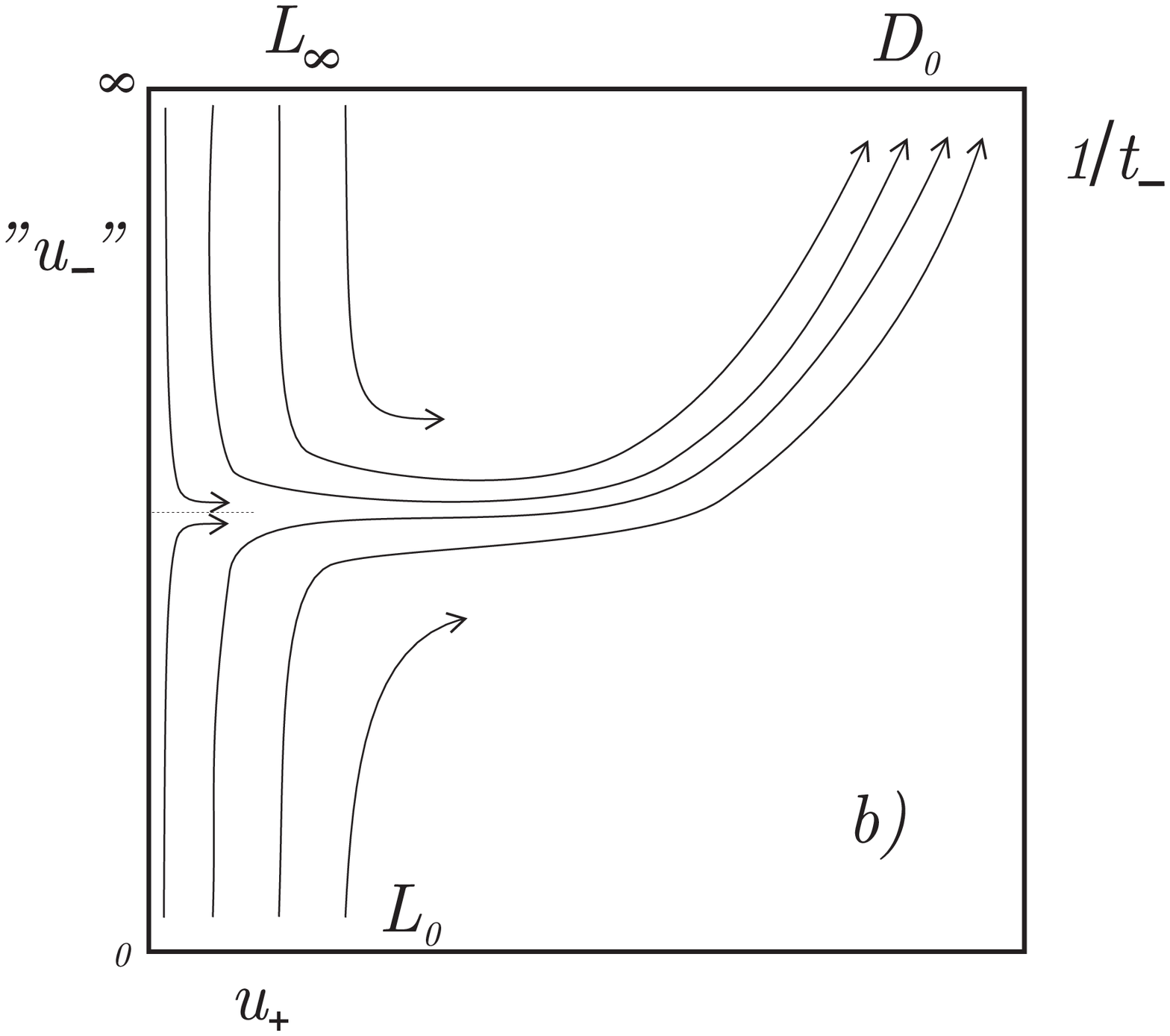}
\caption{ Two possible modifications of the flow diagram that can occur
beyond the approximations of the model, such as the linearization of the
spectrum, the locality of the defect, etc. The flow diagram becomes
dependent on the model parameters in a non universal way. }
\label{fg:RGNL}
\end{figure}

The plot in Fig.~\ref{fg:RGTL} is based on the idealized models, and
as a draft it gives a hint how the known limiting cases could be
matched together. Since the curvature of the electron spectrum is not
universal, different scenarios can occur. The two most apparent
versions of the flow diagram are presented in Figs.~\ref{fg:RGNL}$a$
and~\ref{fg:RGNL}$b$ , but more sophisticated variants can be imagined
due to multidimensionality of the problem, which up to now was hidden
by the linearized spectrum approximation.  In the version presented in
Fig.~\ref{fg:RGNL}$a$, the limiting cases of a weak and a strong
impurity evolve completely independently. In contrast, the RG-flow
presented in Fig.~\ref{fg:RGNL}$b$ corresponds to a scaling from
$L_{0\text{ }}$to $D_{0\text{ }}$, i.e., from a weak impurity
scattering to a limit of two disconnected half-wires, as it has been
assumed in Ref.\cite{EE1D:Kane92}. However, this version of the
RG-flow acquires, in the present discussion, a new essential element.
Namely, the flow trajectory after the first stage where it reaches the $%
L_{f} $ -line dwells at length in its vicinity, and this leads to an
intermediate asymptotic behavior. For a weak enough impurity this
intermediate regime can be very long, and then it determines the low energy
physics in a certain temperature range.

We emphasize that in considerations based on the approximation of a
linearized electron spectrum, RG-trajectories that start at $L_{0\text{ }}$%
end at $L_{f}$. This approximation has been utilized in the mapping of the
problem onto a Coulomb gas theory~\cite{XR:Oreg96,EE1D:Oreg96}. Therefore,
the tunneling density of states, and the Fermi edge singularity exponent,
found in Refs.~\cite{XR:Oreg96,EE1D:Oreg96} correspond to the physics near $%
L_{f}$, and not to a tunneling junction, i.e., not near $D_{0}$.

\section{Summary}

We have studied a single impurity in TL liquids with a repulsive
electron--electron interaction. The problem has been described by a
continuous model with a point-like defect. Two complementary limiting models
corresponding to the weak and the strong impurity potential limits have been
presented on a unified flow diagram depicted in Fig.~\ref{fg:RGTL}. Apart
from the backward scattering amplitude for a weak impurity, and the
tunneling amplitude at the opposite extreme, another parameter
characterizing the ''strength'' of the impurity potential controls the
RG--flow. The unified plot of the flow diagram is rather complex. Because of
many marginal parameters, fixed points of the limiting problems generate
manifolds which on the symbolic two-dimensional plot of the flow diagram we
present as lines. When the potential barrier of the impurity is strong, the
low temperature behavior is controlled by the $D_{0}$-line corresponding to
two disconnected semi--infinite wires. In addition, the flow diagram
contains another attractive fixed line, $L_{f}$, controlling the low
temperature physics when the bare potential of the impurity is weak. The
existence of this line has been established by considering a scattering
Hamiltonian (\ref{rgeq:bih}) with the backward scattering parameter taken to
be very strong, $u_{-}\rightarrow \infty $, while the forward one, $u_{+}$,
was small. This somewhat fictitious case is important due to the fact that
in the TL model only the backward scattering scales to strong coupling while 
$u_{+}$ remains marginal. By mapping this case onto a spin-$1/2$ Heisenberg
chain, it has been shown that the line $L_{\infty }$ describing the limit $%
u_{-}\rightarrow \infty $ is unstable. The existence of the attractive line $%
L_{f}$ at an intermediate value of the parameter $u_{-}$, follows from the
fact that both limiting lines $L_{\infty }$ and $L_{0}$, have proved to be
unstable.

The renormalization of the $u_{-}$-amplitudes resembles the situation of the
overscreened two channel Kondo problem~\cite{RG1D:Nozieres80}. This is not
accidental. At a particular value of the electron--electron interaction
parameter $g=1/2$, the problem of a weak impurity in a TL liquid can be
mapped onto the two channel Kondo model with a specific value of the
longitudinal exchange coupling (for details see the end of Appendix~\ref
{app1}). It is well known that in the overscreened two channel Kondo problem
the limit of infinite exchange interaction is unstable, and there is an
anomalous fixed point at a finite coupling. Since the problem under
consideration, and the overscreened two channel Kondo model are equivalent
at one point, it is natural that the line $u_{-}=\infty $ proves to be
repulsive, as it has been found here. Note, however, that the present
treatment is not restricted to a specific value of the electron--electron
interaction.

The presentation of the phase diagram in the space of parameters
characterizing the impurity potential, helps clarify the difference between
the $D_{0}$- and the $L_{f}$-lines of fixed points--they are located in
different parts of the phase diagram. In the continuous model with a
linearized electron spectrum the two limiting cases of a point-like defect's
strength evolve completely independently. This is constructive for
understanding the low temperature physics in the case of a weak impurity. On
the other hand, an extension beyond the assumption of the model such as the
linearization of the spectrum, the locality of the defect, as well as other
possible mechanisms of a crossover between different regimes, remain open
questions.

A novel line of fixed points can be essential for the scenario of Ref.\cite
{EE1D:Kane92}. This scenario is based on the assumption of scaling from a
weak impurity scattering to a strong barrier. The existence of the $L_{f}$
-line indicates that the situation is more complicated (see the discussion
of Figs.~\ref{fg:RGNL}$a$,~\ref{fg:RGNL}$b$ at the end of Sec.~\ref
{se:RGphaseplane}). We think that the attraction of the weak impurity
problem to $L_{f}$, but not to $D_{0}$, is the reason for the results
obtained by means of the Coulomb gas theory\cite{EE1D:Oreg96,XR:Oreg96}. The
enhancement of the tunneling density of states obtained in this theory
corresponds to decrease of the escape rate of an electron from a defect
center, as a result of multiple backward scattering in combination with the
electron--electron interaction. We believe that the physics of the $L_{f}$%
-line may have a relation to the strengthening of the role of the Friedel
oscillations in the TL model, see e.g., Ref.~\cite{EE1D:Matveev93}.

To conclude, we have developed a continuous model of a single local defect
in a TL liquid. This theory is related to the physics of semiconducting $1D$
quantum wires, and edge states in the quantum Hall effect. We identify a new
attractive fixed point controlling the strong coupling regime of the
backward scattering in the TL model. This novel point may also have
implications for some other related problems, in particular to the theory of
the motion of a quantum particle in a dissipative environment, see Ref.\cite
{RFS:Weiss95} and references therein.

\acknowledgments
We thank Y.~Gefen and D.~E.~Khmelnitskii for useful discussions. A.~F. is
supported by the Barecha Fund Award. The work is supported by the Israel
Academy of Science under the Grant No. 801/94-1 and by German-Israel
Foundation (GIF).

\appendix

\section{~~~}

\label{app1}

In this Appendix the problem of a local impurity in the TL model is mapped
onto a semi-infinite spin-$1/2$ Heisenberg chain, with a magnetic field $%
h\propto u_{-}$ acting on the spin located at origin of the chain. The
mapping onto a spin chain is an appropriate way to study the nature of the
strong coupling regime of the impurity problem. This can be done by
analyzing the stability of the point $h=\infty $. When $h\gg 1$, one can
consider the spin--spin coupling in the spin chain as a small perturbation
in a real space RG analysis. This procedure generates a new operator, which
is not irrelevant, and therefore the fixed point $h=\infty $ proves to be
unstable. Both limits, $h=0$ and $h=\infty $, are repulsive, and
consequently there should be an attractive fixed point at an intermediate
finite value of $h$. Since $h\propto u_{-}$, this analysis yields the
existence of the repulsive $L_{\infty }$-line and the attractive $L_{f}$
-line on the left part of the RG flow diagram depicted in Fig.~\ref{fg:RGTL}.

\subsection{A mapping of impurity scattering in TL liquids onto a spin-$1/2$
semi-infinite chain.}

It is convenient to describe a point-like impurity scattering by a pair of
chiral variables \cite{XR:Affleck94}

\begin{equation}
\begin{array}{lll}
\Theta _{e}(x) & = & \frac{1}{2\sqrt{2}}\left[ \left( \tilde{\phi}(x)+\tilde{%
\phi}(-x)\right) -\left( \phi (x)-\phi (-x)\right) \right] , \\ 
\Theta _{o}(x) & = & \frac{1}{2\sqrt{2}}\left[ \left( \tilde{\phi}(x)-\tilde{%
\phi}(-x)\right) -\left( \phi (x)+\phi (-x)\right) \right] ,
\end{array}
\label{eq:Theta}
\end{equation}
that obey the commutation relations 
\begin{equation}
\begin{array}{lllll}
\left[ \Theta _{e}(x),\Theta _{e}(y)\right] & = & \left[ \Theta
_{o}(x),\Theta _{o}(y)\right] & = & -\frac{i}{4}\mbox {sgn}(x), \\ 
&  & \left[ \Theta _{e}(x),\Theta _{o}(y)\right] & = & -\frac{i}{4}.
\end{array}
\label{eq:thetacom}
\end{equation}
In terms of $\Theta _{e},\Theta _{o}$ the Hamiltonian $H=H_{TL}+H_{sc}$, see
Eqs. (\ref{rgeq:bfh}) and (\ref{rgeq:bih}), can be rewritten in the form 
\[
H=H_{e}+H_{o}\ , 
\]
\begin{equation}
H_{o}=\frac{v_{F}}{g}\int dx\left[ \left( \frac{\partial \Theta _{o}}{%
\partial x}\right) ^{2}-u_{-}\frac{g}{\pi \eta }\cos \left( \beta \sqrt{2}%
\Theta _{o}(x)\right) \delta (x)\right] ,  \label{eq:LAH}
\end{equation}
\[
H_{e}=\frac{v_{F}}{g}\int dx\left[ \left( \frac{\partial \Theta _{e}}{%
\partial x}\right) ^{2}-u_{+}\frac{\beta g}{\sqrt{2}\pi }\frac{\partial
\Theta _{e}(x)}{\partial x}\delta (x)\right] . 
\]
Although $\Theta _{e}$ and $\Theta _{o}$ do not commute, the Hamiltonian $H$
is divided into even and odd parts, $H_{e}$ and $H_{o}$, respectively,
because the even part contains only derivatives of $\Theta _{e}$. For small
bare scattering amplitudes $\delta _{\pm }$, and a not too strong
electron--electron interaction, the linearization of the spectrum of the
electrons used in the TL model is valid. The decoupling of forward and
backward scattering holds under this approximation. For simplicity we omit
the phase $\varphi _{u}$, and the $u_{a}$-term related to time reversal
asymmetry, in $H_{o}$.

We now show that the odd part $H_{o}$ is effectively equivalent to the
Hamiltonian of a semi-infinite spin-$1/2$ antiferromagnetic chain with
anisotropy $\overline{\gamma }$: 
\begin{equation}
H_{s}=\frac{J}{2}\sum_{n=0}^{\infty }\left(
S_{n}^{+}S_{n+1}^{-}+S_{n}^{-}S_{n+1}^{+}\right) +\overline{\gamma }%
J\sum_{n=0}^{\infty }\left( S_{n}^{+}S_{n}^{-}-\frac{1}{2}\right) \left(
S_{n+1}^{+}S_{n+1}^{-}-\frac{1}{2}\right) -hJ\left(
S_{0}^{-}+S_{0}^{+}\right) .  \label{eq:spin}
\end{equation}
A Hamiltonian of this type, with $\overline{\gamma }=0$, has been introduced
by Guinea \cite{EE1D:Guinea85} for the description of a quantum particle
interacting with a dissipative environment, at a particular value of the
friction coefficient. It has also been used to discuss the transmission
through barriers in TL liquids\cite{EE1D:Kane92a}, for a given value of the
electron--electron interaction $g=1/2$. Here we introduce the $\overline{%
\gamma }$-term in order not to be limited to a particular value of the
electron--electron interaction. Note, that in $H_{s}$, unlike the standard
tight binding model (see, e.g., Ref. \cite{RG1D:Eggert92}), the defect is
located on a single site $j=0$, and therefore it has no internal structure.
This is consistent with our purpose to study a point-like impurity.

To show the equivalence of $H_{o}$ and $H_{s}$, one should perform a
sequence of transformations. After applying the inverse of the Jordan-Wigner
transformation\cite{RFS:Fradkin91,RG1D:Luther75} 
\[
S_{n}^{+}=c^{\dagger }(n)e^{i\pi \sum_{j=0}^{n-1}c_{j}^{\dagger }c_{j}},\
S_{n}^{-}=e^{-i\pi \sum_{j=0}^{n-1}c_{j}^{\dagger }c_{j}}c(n), 
\]
$H_{s}$ transforms into $H_{c}$, where 
\begin{equation}
H_{c}=\frac{J}{2}\sum_{j=0}^{\infty }c_{j}^{\dagger }c_{j+1}+h.c.+\overline{
\gamma }J\sum_{j=0}^{\infty }\left( n_{j}-\frac{1}{2}\right) \left( n_{j+1}- 
\frac{1}{2}\right) -hJ\left( c_{0}^{\dagger }+c_{0}\right) .
\label{eq:spinfermion}
\end{equation}
In the absence of the local magnetic field $h$, the average of $%
S^{z}=\sum_{i}S_{i}^{z}$ over the ground state $\left| G\right\rangle $ of
the spin chain is equal to zero. Since

\[
S^{z}=\sum_{j=0}^{N}c_{j}^{\dagger }c_{j}-N/2=N_{F}-N/2, 
\]
the sector $\left\langle G\right| S^{z}\left| G\right\rangle =0$ corresponds
to half filling. The local operator $hJS_{0}^{x}$ cannot change the bulk
properties of the chain. Therefore, it will be assumed that the fermion
system $H_{c}$ of Eq.(\ref{eq:spinfermion})\ is at half filling. The
continuum limit of $H_{c}$ (e.g., see Ref. \cite{RFS:Fradkin91}) corresponds
to the effective Hamiltonian $H_{c}^{cont}=H_{0}+H_{int}+H_{h}$, where 
\begin{equation}
\begin{array}{lll}
H_{0} & = & iv_{F}\int_{0}^{\infty }dx\left( L^{\dagger }(x)\partial
_{x}L(x)-R^{\dagger }(x)\partial _{x}R(x)\right) , \\ 
H_{int} & = & v_{F}\bar{\gamma}\int_{0}^{\infty }dx\left( \bar{\rho}%
_{L}{}^{2}+\bar{\rho}_{R}{}^{2}+4\bar{\rho}_{R}\bar{\rho}_{L}\right) -2\bar{%
\gamma}v_{F}\int_{0}^{\infty }dx\left[ \left( R^{\dagger }(x)L(x)\right)
^{2}+\left( L^{\dagger }(x)R(x)\right) ^{2}\right] , \\ 
H_{h} & = & v_{F}\frac{1}{\sqrt{\eta }}h\left( R(x=0)+L(x=0)+R^{\dagger
}(x=0)+L^{\dagger }(x=0)\right) .
\end{array}
\label{eq:rl}
\end{equation}
Here $v_{F}=J\eta $, where $\eta $ is the lattice spacing, and the operators 
$L$ and $R$ represent left and right movers on a semi-infinite line. The
remnant of the discrete structure of the chain is the last term in $H_{int},$
which corresponds to the Umklapp processes at half filling\cite
{RG1D:Haldane82,RG1D:Haldane80}. This Umklapp term scales to zero for $%
\left| \bar{\gamma}\right| <1$, and also renormalizes the electron--electron
interaction with a small momentum transfer. However, at small $\bar{\gamma}$
the latter effect is negligible, and for the low energy description one may
substitute $H_{int}$ by the effective term 
\begin{equation}
H_{int}^{\prime }=v_{F}\overline{\gamma }\int_{0}^{\infty }\left( \bar{\rho}%
_{L}{}^{2}+\bar{\rho}_{R}{}^{2}+4\bar{\rho}_{R}\bar{\rho}_{L}\right) dx.
\label{eq:halfint}
\end{equation}
The last step, which needs to be carried out in order to establish the
equivalence between $H_{o}$ and $H_{s}$, is to unfold the semi-infinite line
with left and right movers into a full line with a single chiral bosonic
field. This yields: 
\begin{equation}
\begin{array}{lll}
H_{chiral} & = & \pi v_{F}\int_{-\infty }^{\infty }dx\left[ \rho
_{ch}^{2}(x)+\overline{\gamma }/\pi \left( \rho _{ch}^{2}(x)+2\rho
_{ch}(x)\rho _{ch}(-x)\right) \right] \\ 
&  & -\frac{4v_{F}}{\sqrt{2\pi }\eta }h\cos \left( \sqrt{4\pi }\Theta
_{ch}(x=0)\right) ,
\end{array}
\label{eq:chiralL}
\end{equation}
where $d\Theta _{ch}(x)/dx=\sqrt{\pi }\rho _{ch}(x).$ Diagonalizing the
quadratic part of the Hamiltonian, we find: 
\begin{equation}
\overline{H}_{chiral}=\frac{v_{F}}{g_{ch}}\int_{-\infty }^{\infty }dx\left[
\pi \overline{\rho }_{ch}^{2}(x)-\frac{4g_{ch}}{\sqrt{2\pi }\eta }h\cos
\left( \beta _{ch}\bar{\Theta}_{ch}(x)\right) \delta (x)\right] ,
\label{eq:chiralLbar}
\end{equation}
where $\beta _{ch}=\sqrt{4\pi }\exp $ $\chi $, $\chi =\frac{1}{2}$arctanh$%
\left( 2\overline{\gamma }/\left( \pi +\overline{\gamma }\right) \right) $,
and $g_{ch}\approx \left( 1+\overline{\gamma }/\pi \right) ^{-1}$ . Notice,
the important role of the $\overline{\gamma }$-term -- it modifies $\beta
_{ch}$ inside the cosine term.

Thus, as a result of the sequence of transformations 
\begin{equation}
H_{s}\Longrightarrow H_{c}\Longrightarrow H_{c}^{cont}\Longrightarrow
H_{chiral}\Longrightarrow \overline{H}_{chiral}\Longrightarrow H_{o},
\label{eq:sequence}
\end{equation}
we obtain that the Hamiltonians $H_{s}$ and $H_{o}$ are equivalent when $%
\beta _{ch}=\sqrt{2}\beta $ and$\;h=u_{-}/\sqrt{8\pi }$. Due to the $%
\overline{\gamma }$-term, this equivalence is extended here to a finite
interval of the electron--electron interaction.

\subsection{A scaling analysis of the spin chain model.}

We will use the equivalence of $H_{o}$ and $H_{s}$ to analyze the stability
of the fixed line at $u_{-}=\infty $ and small $u_{+}$. A simple variant of
the Nozi\`{e}res and Blandin approach\cite{RG1D:Nozieres80} in their
analysis of the two channel Kondo problem will be considered. Following this
approach, we will assume that $h\gg 1,$ and check whether the fixed point $%
h=\infty $ is a stable one.

In the presence of a strong magnetic field $h\gg 1$, the spin at site $0$ is
oriented along the direction opposite to the magnetic field. Its coupling to
the nearest neighbor at the lattice site $1,$ can be treated as a
perturbation. The reduced Hamiltonian that includes only sites $0$ and $1$
is given by 
\begin{equation}
H_{s}^{01}=\frac{J}{2}\left( S_{0}^{+}S_{1}^{-}+S_{0}^{-}S_{1}^{+}\right) +%
\overline{\gamma }J\left( S_{0}^{+}S_{0}^{-}-\frac{1}{2}\right) \left(
S_{1}^{+}S_{1}^{-}-\frac{1}{2}\right) -hJ\left( S_{0}^{-}+S_{0}^{+}\right) .
\label{eq:Jper1}
\end{equation}
After performing the permutation $x\rightarrow z$, $y\rightarrow x$, and $%
z\rightarrow y$ we obtain the Hamiltonian 
\begin{equation}
H_{s}^{01}=JS_{0}^{z}S_{1}^{z}+\frac{J}{4}\left( 1-\bar{\gamma}\right)
\left[ S_{0}^{+}S_{1}^{+}+S_{0}^{-}S_{1}^{-}\right] +\frac{J}{4}\left( 1+%
\bar{\gamma}\right) \left[ S_{0}^{-}S_{1}^{+}+S_{0}^{+}S_{1}^{-}\right]
-2hJS_{0}^{z}.  \label{eq:Jper2}
\end{equation}
For $h\gg 1$ the spin at site $0$ is in the state $\left| 0\uparrow
\right\rangle $. Up to the first order in $J$ we have 
\[
\left\langle 1\downarrow \right| \left\langle 0\uparrow \right|
H_{s}^{01}\left| 0\uparrow \right\rangle \left| 1\downarrow \right\rangle
=-J/4,\qquad \left\langle 1\uparrow \right| \left\langle 0\uparrow \right|
H_{s}^{01}\left| 0\uparrow \right\rangle \left| 1\uparrow \right\rangle
=J/4. 
\]
This means that the spin at site $1$ is under the action of an effective
magnetic field $\tilde{h}=-1/4$. Under the assumption that $h\gg 1$, the
higher orders in the perturbation theory give small corrections of the order
of $h^{-1}$. As a result of this renormalization procedure step, we arrive
at a problem equivalent to the initial one: a semi-infinite spin-$\frac{1}{2}
$ Heisenberg chain, with a local magnetic field acting on the site at origin
of the chain (now it will be site $1$).

If one assumes that the discussed fixed point is such that the local
magnetic field at the origin of the spin chain flows to infinity, then the
renormalization procedure generates a relevant operator that makes this
process nonconvergent. This is in contradiction with the initial assumption
that $h=\infty $ is a stable fixed point. We have to conclude that $h=\infty 
$ is a repulsive fixed point. Note, that this conclusion holds for a finite
interval of the parameter $\left| \bar{\gamma}\right| \lesssim 1$, because $%
\bar{\gamma}$ does not radically influence the effective magnetic field
acting on site $1$. One can see this from the second order corrections to
the energy of the anti-parallel and parallel spin configurations, that are $%
-J\frac{\left( 1\pm \bar{\gamma}\right) ^{2}}{32}h^{-1}$.

It is not accidental that the present discussion resembles the analysis of
the overscreened two channel Kondo problem\cite{RG1D:Nozieres80}. Indeed,
the spin chain model, in the absence of the $\overline{\gamma }$-term, is
equivalent to the TL-impurity problem at a particular value of an
electron--electron interaction parameter $g=1/2$ \cite
{EE1D:Guinea85,EE1D:Kane92a}. The latter problem can in turn be reduced to a
resonant level model\cite{EE1D:Matveev95}. Moreover, the overscreened two
channel Kondo model at a specific value of the longitudinal exchange
coupling is equivalent to the same resonant level model\cite{KE&AMM:Emery92}%
. Thus, the spin chain model and the Kondo model are equivalent at one
point. On the other hand, it is well known that in the overscreened two
channel Kondo problem the limit of infinite exchange interaction is
unstable, and there is an anomalous fixed point at a finite coupling\cite
{RG1D:Nozieres80}. This property is preserved in the presence of the spin
exchange anisotropy, which is irrelevant\cite{RG1D:Affleck92}. Since the
spin chain model and the two channel Kondo model are equivalent at one
point, it is natural that we have found that the point $h=\infty $ is
repulsive. Since $h\propto u_{-}$, it follows from this analysis that the
line $u_{-}\gg 1$ is a repulsive fixed line for the problem of impurity
scattering in a TL liquid. The present treatment is not restricted to the
special point of a TL liquid with $g=1/2$. This has been accomplished by the 
$\overline{\gamma }\neq 0$ term in the spin chain model.

%\bibliographystyle{prsty}
%\bibliography{/pusers1/fnoreg/ref/library}

\begin{thebibliography}{10}
\bibitem{RG1D:Weaklink}
Sometimes it is also called 'a weak link junction' or 'open boundary
  conditions'.
\end{thebibliography}

\end{document}